\documentclass[nobibnotes,preprintnumbers,aps,prl,superscriptaddress,showpacs,twocolumn,twoside,sort&compress]{revtex4}				 %
\usepackage{graphicx, float}																												 %
\usepackage{titlesec}																											     %
\usepackage{fancyheadings}																							 				 %
\begin{document}																													 %
\preprint{{Vol.XXX (201X) ~~~~~~~~~~~~~~~~~~~~~~~~~~~~~~~~~~~~~~~~~~~~~~~~~~~~ {\it CSMAG`16}										  ~~~~~~~~~~~~~~~~~~~~~~~~~~~~~~~~~~~~~~~~~~~~~~~~~~~~~~~~~~~~ No.X~~~~}}																 %
\vspace*{-0.3cm}																													 %
\preprint{\rule{\textwidth}{0.5pt}}																											 \vspace*{0.3cm}																														 %

\title{Magnetization curves of geometrically frustrated exchange-biased FM/AFM bilayers }

\author{M. Pankratova}
\thanks{corresponding author; e-mail: pankratova$\_$mari@mail.ru}
\affiliation{Institute of Physics, Faculty of Sciences, P. J. \v{S}af\'{a}rik University, Park Angelinum 9, 041 54 Ko\v{s}ice, Slovakia}
\author{M. \v{Z}ukovi\v{c}}
\affiliation{Institute of Physics, Faculty of Sciences, P. J. \v{S}af\'{a}rik University, Park Angelinum 9, 041 54 Ko\v{s}ice, Slovakia}

\begin{abstract}
We consider a ferromagnetic/antiferromagnetic bilayer on a triangular lattice in
the framework of the planar Heisenberg model. The impact of the geometrical
frustration in this system on the magnetization curves and the exchange bias phenomenon is studied.
The magnetization curves and the phase diagram for such systems are obtained.
We observe  horizontal plateaus and a split of the hysteresis loop on
the magnetization curves. It is shown that the shift of the hysteresis loop (exchange bias) occurs
for the systems with a magnetically hard antiferromagnet.
\end{abstract}


\maketitle

%
\section{1. Introduction}
\vskip -0.20cm
In this paper we study the exchange bias phenomenon in the ferromagnetic/ antiferromagnetic bilayer (FM/AFM) with
the geometrical frustration. The exchange bias phenomenon consists in the shift of the magnetic hysteresis loop $M(H)$
along the external magnetic field axis $H$ \cite{1}. Moreover, some experimental studies show the asymmetry of a
magnetization curve, the appearance of horizontal plateaus, and the split of hysteresis loops. These effects are widely
studied theoretically and experimentally  in layered FM/AFM systems
but still have no comprehensive explanation. Despite a large number of works, the influence of the geometrical frustration
 in the bilayer system on the exchange bias has been little studied yet  \cite{frust}, \cite{frust-exp}.
The geometrical frustration appears when the minimum of the system energy does not correspond the minimum
of all local interactions. The triangular lattice with the AFM interaction between
each pair of spins is a simplest example of the geometrically frustrated system. In this system the frustration appears because
 of incompatibility between the local interactions and the lattice geometry.
In this paper the FM/AFM bilayers on the triangular lattice
are studied in the framework of the planar Heisenberg model with periodic boundary conditions.
The outline of the paper is as follows. In Section 2 we introduce a layered system
made of one FM and one AFM monolayers on a triangular lattice. Section 3  is devoted
to the study of the FM/AFM structure with fixed (so called frozen) AFM magnetic moments.
The paper is completed by the Concluding remarks.

\section{2. FM/AFM bilayer on triangular lattice}
\label{mod}
In this section we consider a FM/AFM bilayer made of
two monolayers on the triangular lattice. The interaction through the FM/AFM
interface $J_1>0$ is considered to be ferromagnetic. The exchange interactions in the FM
and AFM films are given by the parameters $J>0$ and $J_{0}<0$, respectively. We assume a
strong easy-plane anisotropy both in the AFM (layer A) and FM (layer B) layers
and an additional single-ion anisotropy $\beta_i$, $i=A,B$ in the easy planes of the FM and AFM
subsystems. We consider different values of the magnetic anisotropy $\beta_i$ for
the AFM and FM planes $\beta_{A} \ne \beta_{B}$. It is assumed
that the external magnetic field $H$ is directed along the easy axis.The magnetic states of the
magnetization vectors in this bilayer are given by the rotational angles $\varphi_i$ of spins in
the easy plane. The magnetic energy of the system is given by:
$$
E=-J_0 \sum_{i,j \in A} \cos(\varphi_{i} - \varphi_{j})-J \sum_{i,j \in B}\cos(\varphi_{i}- \varphi_{j})
$$
\vskip -0.4 cm
$$
- J_1 \sum_{i \in A, j \in B} \cos(\varphi_{i}- \varphi_{ j})  -H \sum_{i \in A,B} \cos(\varphi_{i})
$$
\vskip -0.3 cm
\begin{equation}
\label{energy}
 - \frac{\beta_{B}}{2} \sum_{i \in B} \cos^2(\varphi_{i})
- \frac{\beta_{A}}{2} \sum_{i \in A} \cos^2(\varphi_{i}),
\end{equation}

where the first three summations run over all nearest neighbours and the next three over the spins in the respective layers. Assuming spin uniformity in the respective sublattices the possible equilibrium states are given by the equations:
\begin{equation}
\label{e-varphi}
\partial E/ \partial \varphi_{k,l}=0, \quad (k = 1..3, l=A,B).
\end{equation}
where indexes $l$ and $k$ correspond to the  planes (FM or AFM) and to the sublattices in these planes respectively.
The solutions of these equations are the parallel structures
($\varphi_{k,l}=0,\pi$), the non-collinear structures  ($\varphi_{k,l} \ne 0, \pi$),
and the antiparallel structures which correspond to the horizontal plateaus
($\varphi_{2,A}=0, \varphi_{2,B}=\pi, \varphi_{k,l}=\pi$), where $k=1,3, l=A,B$, so all magnetic moments are laying along the field
exept one in the layer A that has an opposite direction. The second type of horizontal plateaus cooresponds to a
$\varphi_{2,l}=0, \varphi_{k,l}=\pi$, where $k=1,3, l=A,B$. 

The transition from the collinear phase to the canted phase corresponds to the
bifurcation of the solutions $\varphi_{k,l}=0,\pi$. In the neighborhood of the bifurcation
point, there are canted solutions of Eq. (\ref{e-varphi})
which are infinitesimally close to the collinear states. To find this point,
we linearize these equations with respect to the angles $\varphi_{k,l}$ and look for the
nonzero solutions of the linearized equations. The analysis of the stability of the
$\varphi_{k,l}=\pi$ phase can be done in a similar way.

\begin{figure}[h!]
\includegraphics[width=1.0\columnwidth, height=5cm]{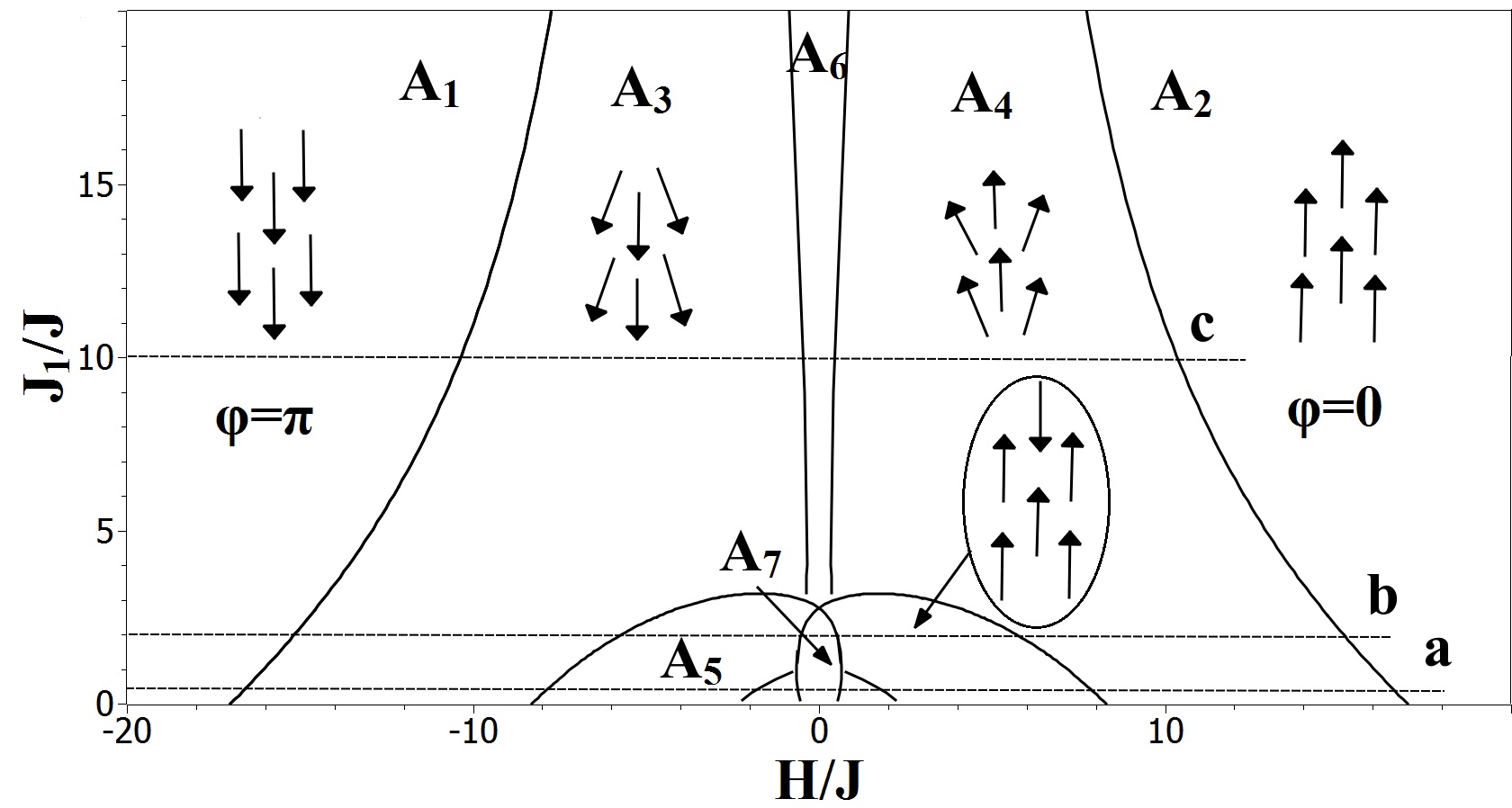}
\caption{Phase diagram of the FM/AFM bilayer on the triangular lattice for the values of the parameters $ J_0/J=-2, \beta_A/J=1, \beta_B/J=0.5$.
The dashed lines correspond to the magnetization curves in Fig.\ref{fig2-P1-20} a-c.}
\label{fig1-P1-20}
\end{figure}

\begin{figure}[h!]
\includegraphics[width=0.8\columnwidth,height=4.5cm]{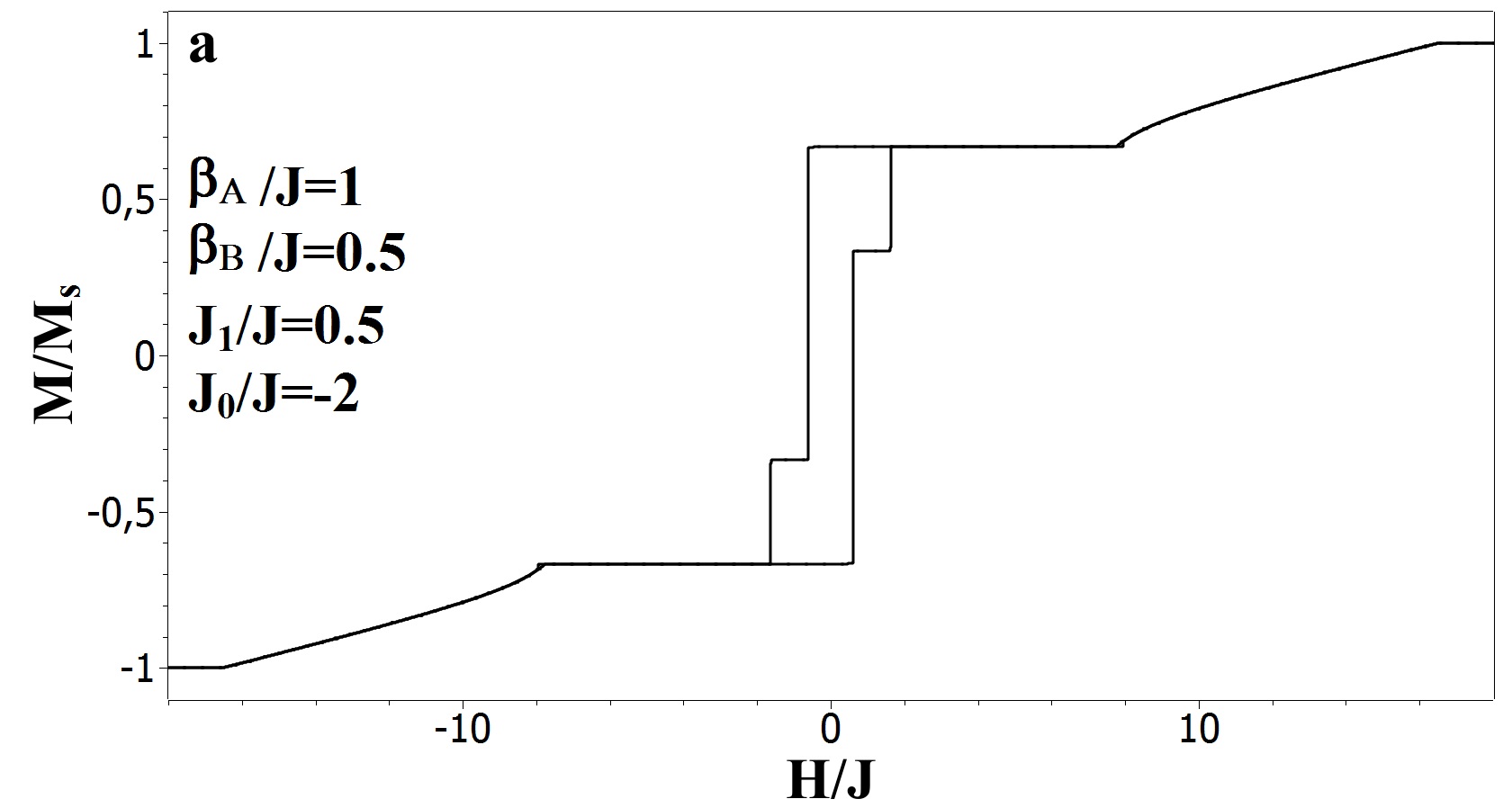}
\includegraphics[width=0.8\columnwidth,height=4.5cm]{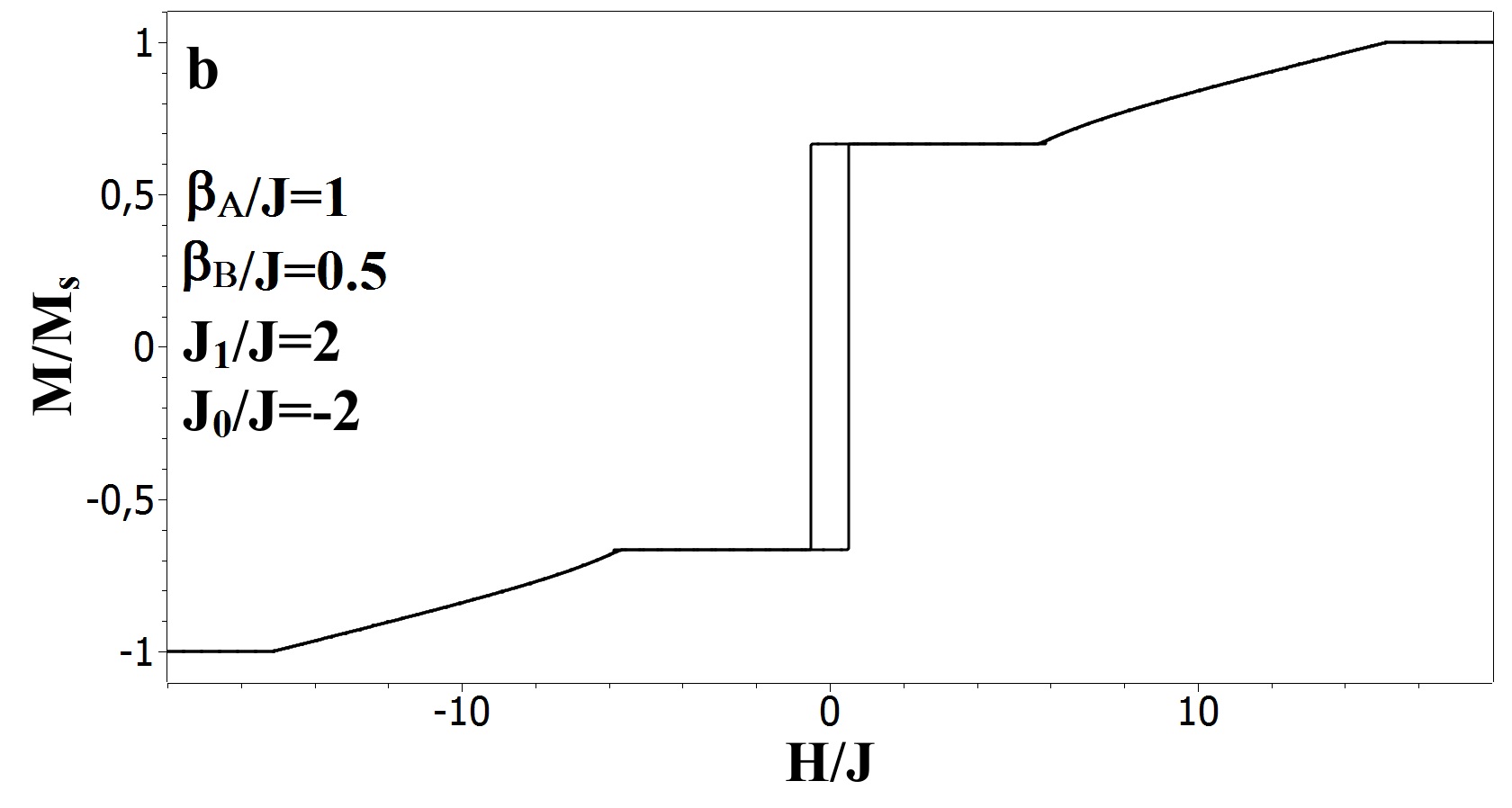}
\includegraphics[width=0.8\columnwidth,height=4.5cm]{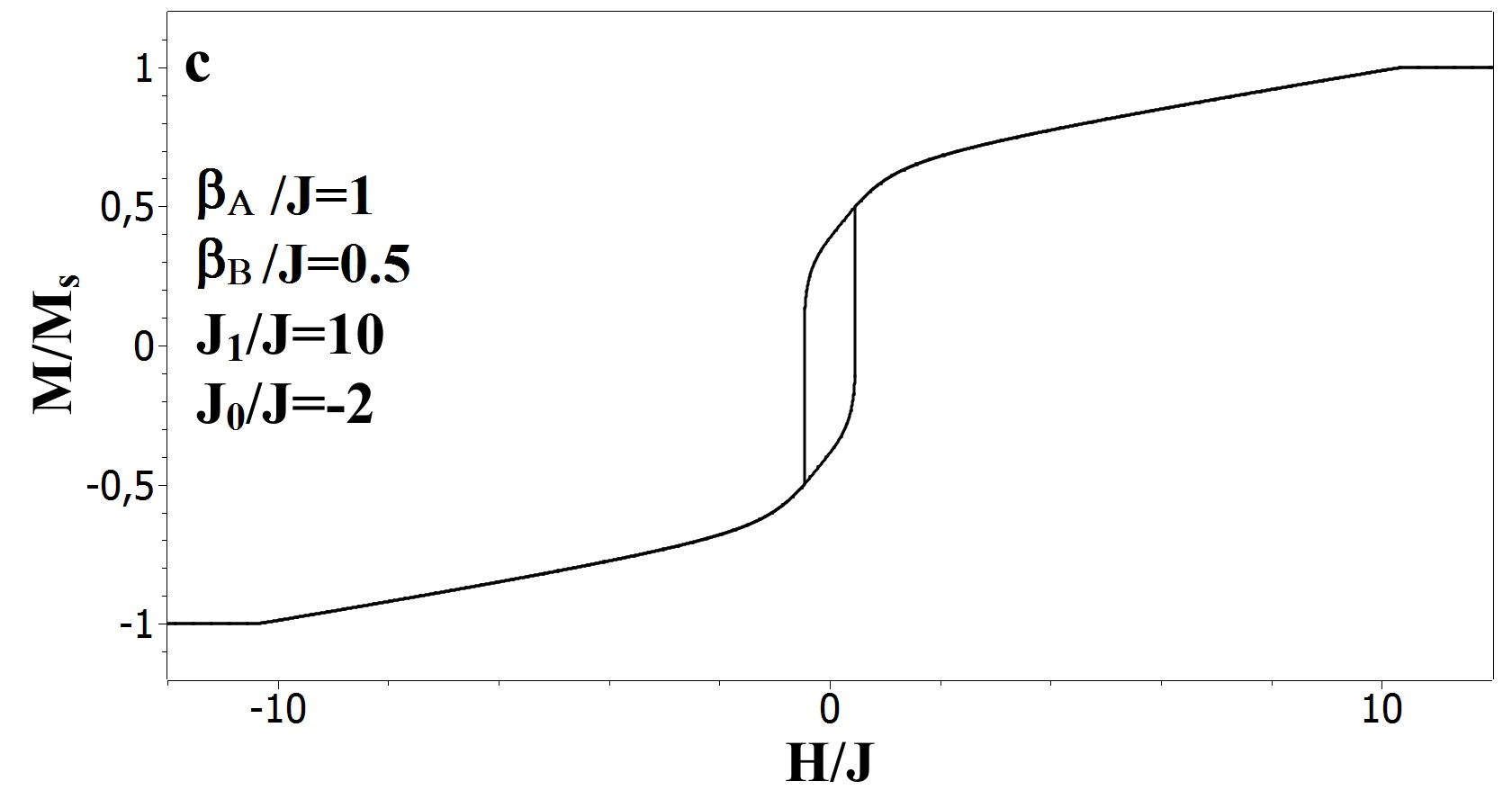}
\includegraphics[width=0.8\columnwidth,height=4.4cm]{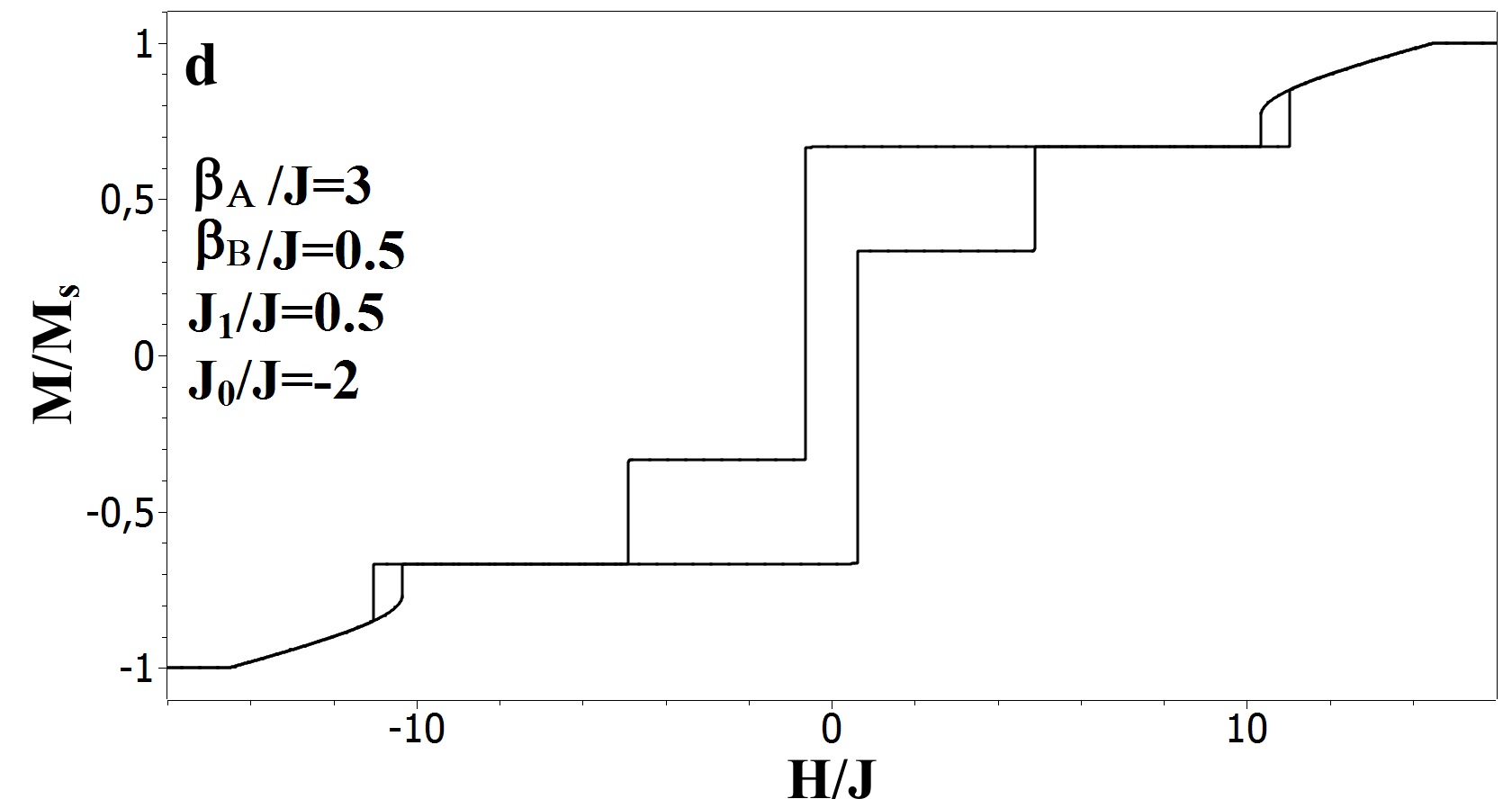}
\caption{ The hysteresis loops of the FM/AFM bilayer for different values of $J_1$ (a-c) and $\beta_A$ (a,d).}
\label{fig2-P1-20}
\end{figure}

The areas of the existence of the identified structures are given in
Fig.\ref{fig1-P1-20}. The arrows show the direction of magnetization.
 The parameter regions A1 and A2 in Fig.\ref{fig1-P1-20}
 correspond to the parallel states in the FM/AFM bilayer, A3 and A4 to the non-collinear structures, and A5 to the existence of the antiparallel states.
The small area $A_7$ corresponds to the coexistence of  horizontal plateaus (see hystersis loop in the Fig.\ref{fig2-P1-20}a,b,d).
The area  $A_6$ is the domain of the existence of the hysteresis loop in
the magnetization curves.  In the areas $A_1, A_2$ the external magnetic field is big enough to
reverse both FM and AFM and the magnetization vectors in both layers have the same direction.
On the contrary, the direction of magnetic moments of the layer A and layer B in the other areas can be different as for example in the horizontal plateau, described above.
The lines for the boundaries of the parallel and antiparallel states
were obtained analytically from Eq. (\ref{e-varphi}), while the other curves could only be obtained
numerically.

The total magnetization of FM/AFM bilayer is given by the formula:
$
M=\sum_{i \in A,B}{\cos{\varphi_i}}
$. The hysteresis loops were obtained for different values of $J_1$ (Fig.\ref{fig2-P1-20}a-c)  and $\beta_A$ (Fig.\ref{fig2-P1-20}a,d).
%

\section{3. The case of frozen AFM}
\label{frozen}
 In this section the AFM layer in the FM/AFM bilayer is considered to be magnetically hard,
i.e., for the magnetic fields that are less than the spin-flop transition,
its magnetic structure is fixed during the entire magnetization reversal\cite{kiwi}. One can obtain the frozen AFM layer as a surface of a staked AFM with large anisotropy. In this AFM stack the structure of each layer is ferrimagnetic ($\uparrow \downarrow \uparrow$), with magnetic moments in each layer
having opposite direction with their nearest neighbors in the neighbouring layers and thus the total magnetization of the AFM stack is zero.
We consider a case of an uncompensated FM/AFM interface. In particular, the collinear structure is considered with two magnetic moments
 in the triangular plaquette laying opposite to the external magnetic field and one magnetic moment is laying along the field.
The uncompensated AFM interface can appear in the following way. Because of the frustration, a noncollinear structure with
zero magnetization appears in the AFM layer in zero magnetic field. Due to the interaction with the FM layer the spins in the
 AFM layer deviate from their original positions and thus form an uncompensated AFM interface \cite{frust}.

\begin{figure}[H]
\includegraphics[width=0.950\columnwidth,height=5cm]{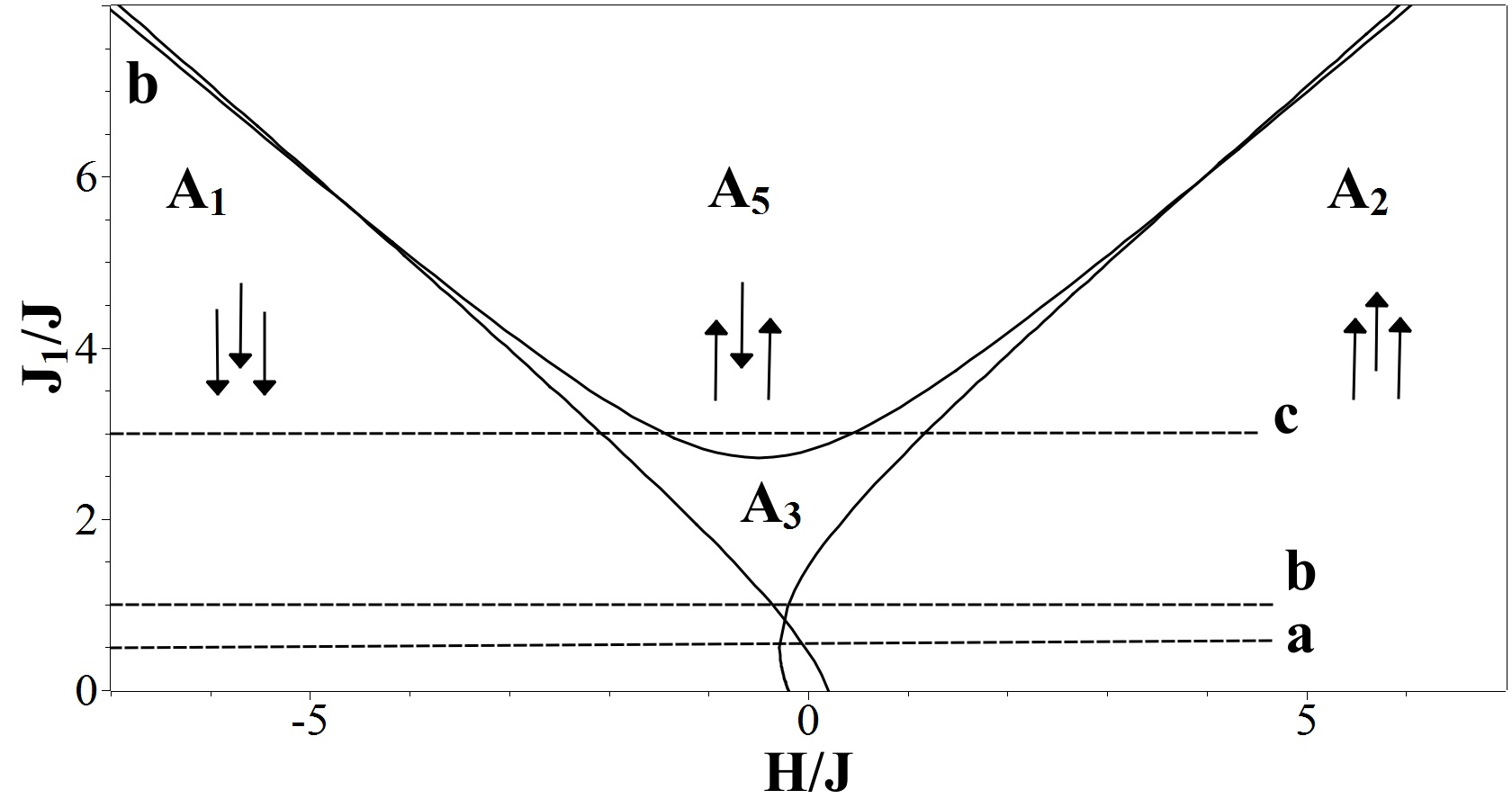}
\caption{ Phase diagram of the FM/AFM bilayer on the triangular lattice.
The case of a fixed AFM for $\beta_B/J=2$.The dashed lines correspond to the magnetization curves in Fig.4.}
\label{fig3-P1-20}
\end{figure}
\begin{figure}[H]
\includegraphics[width=0.8\columnwidth,height=4.5cm]{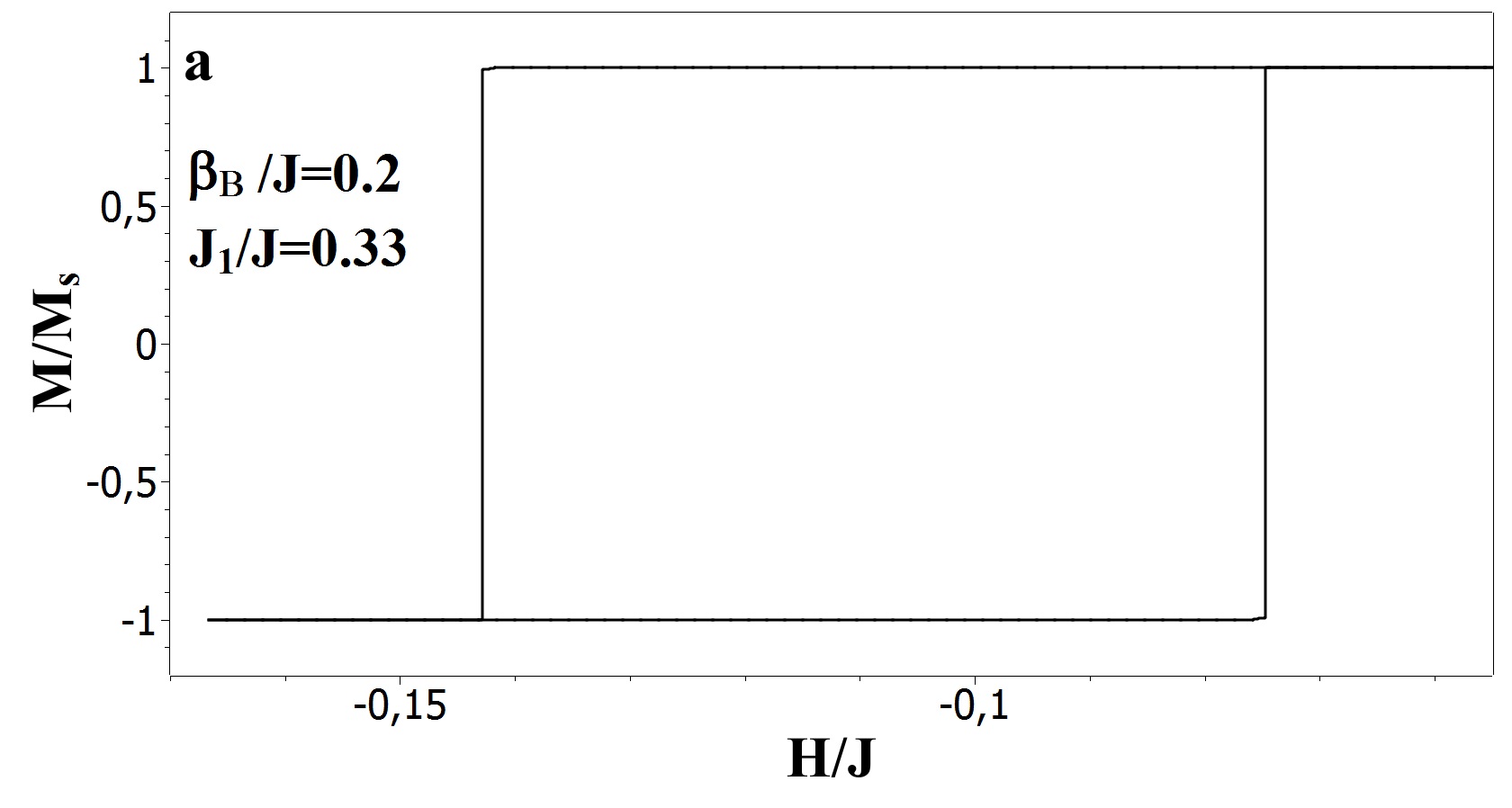}
\includegraphics[width=0.8\columnwidth,height=4.5cm]{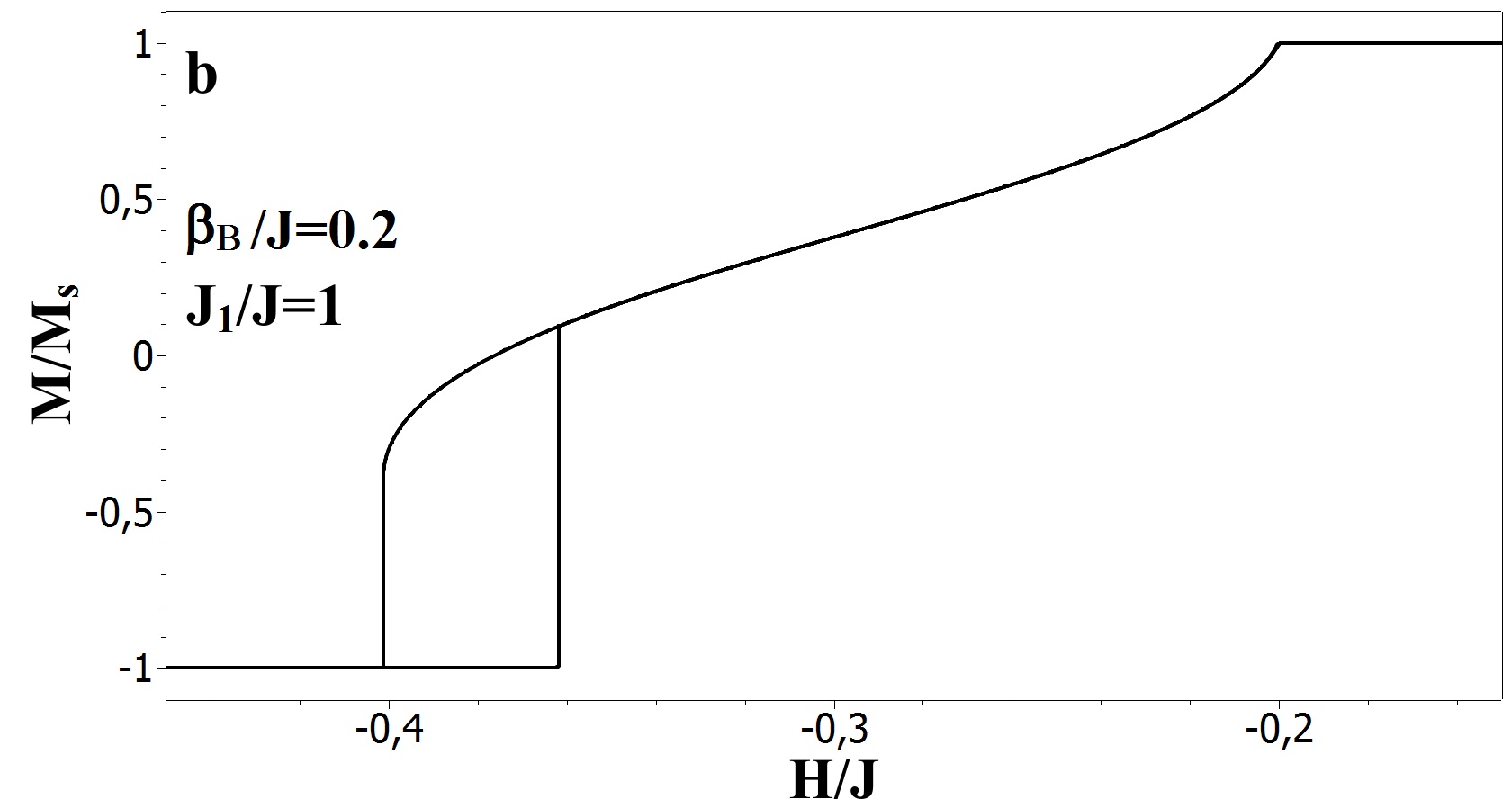}
\includegraphics[width=0.8\columnwidth,height=4.5cm]{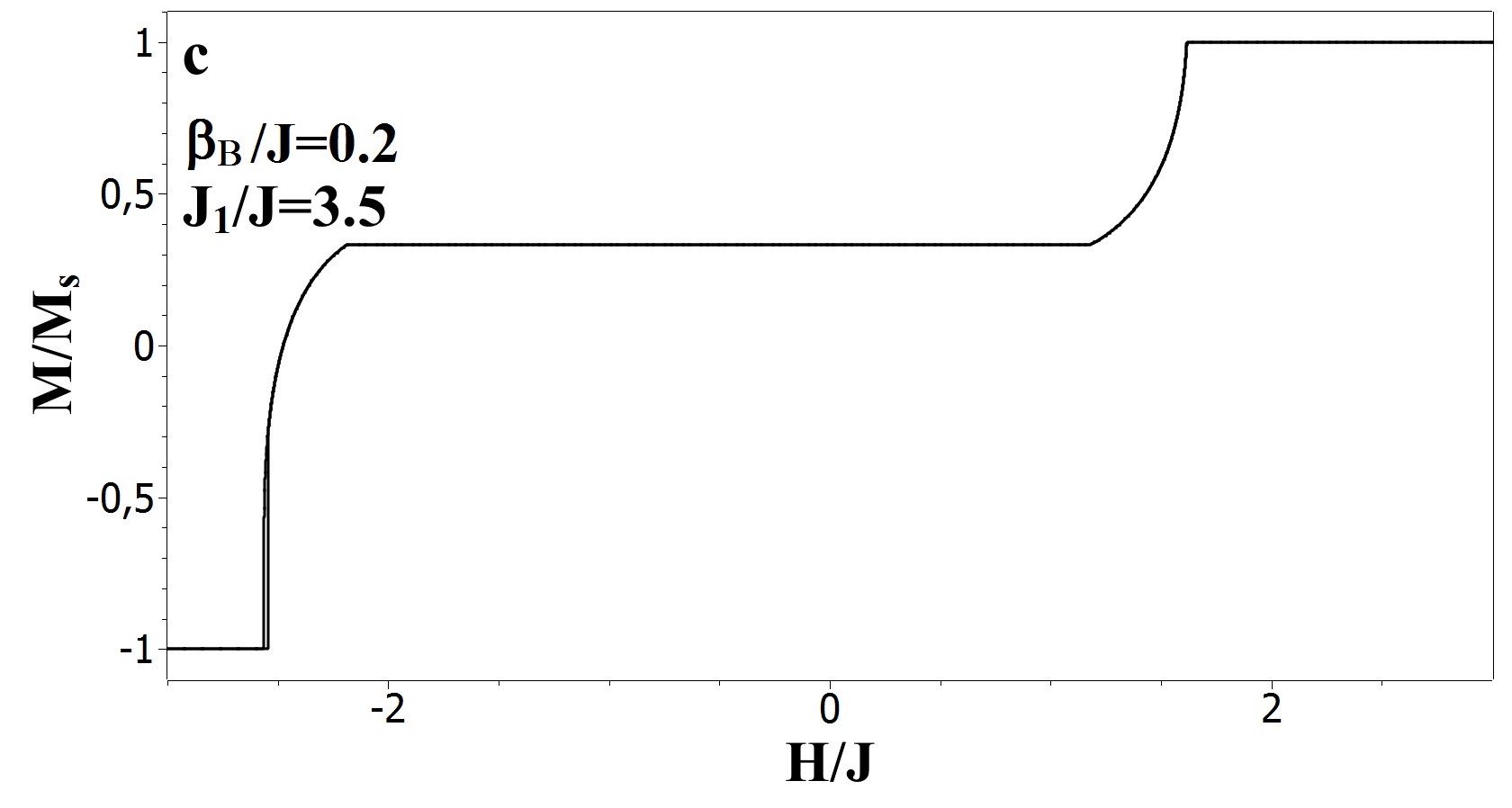}
\caption{ The hysteresis loops of the FM/AFM bilayer for different values of $J_1$.}
\label{fig4-P1-20}
\end{figure}

 The areas of existence of the paralell $A_1, A_2$,
 non-collinear $A_3$, and antiparallel $A_5$ phases (Fig.\ref{fig3-P1-20}) and the hysteresis loops (Fig.\ref{fig4-P1-20}) were obtained
in a way similar to that from the previous section. It is shown that in this model the shifted  magnetization curve can be asymmetric (Fig.\ref{fig4-P1-20} b,c) and
has horizontal plateaus (Fig.\ref{fig3-P1-20} c). In this section the $M(H)$ curves correspond to
the FM layer, while the magnetic moments of AFM are fixed and do not contribute to the magnetization.

\section{Conclusions}
In the framework of the planar Heisenberg model the magnetization curves of the FM/AFM bilayer
on a triangular lattice are studied.
We have considered the cases of the non-frozen and frozen AFM. The hysteresis loops have been
obtained for different values of the exchange interaction and the magnetic anisotropy.
The exchange bias is observed in the case of the frozen AFM. Horizontal plateaus
 and the hysteresis loops are observed for both frozen and non-frozen AFM cases. Phase diagrams
 have been calculated for selected values of the parameters.

\newpage
\vfill
\section{Acknowledgement}
This work  was supported  by the Scientific Grant Agency of Ministry of Education of Slovak Republic 
(Grant No. 1/0331/15) and by the National Scholarhip Programme of the Slovak Republic.

\vfill \clearpage
\end{document}